\documentclass[10pt,aps,prb,superscriptaddress,twocolumn,floatfix]{revtex4-2}

\usepackage{graphicx}
\usepackage{xcolor}
\usepackage{url} 
\usepackage{hyperref} 
\usepackage{dcolumn}
\usepackage{bm}
\usepackage{multirow}
\usepackage[symbol]{footmisc}
\usepackage{amsmath}
\usepackage{amssymb}
\usepackage{amsfonts}
\usepackage{subfigure}
\usepackage{url}
\usepackage{soul}

\newcommand{\czm}{Co$_8$Zn$_8$Mn$_4$}

\begin{document}

\title{Competing anisotropies in the chiral cubic magnet Co$_8$Zn$_8$Mn$_4$ unveiled by resonant x-ray magnetic scattering}

\author{Victor Ukleev}
\email{victor.ukleev@helmholtz-berlin.de}
\affiliation{Helmholtz-Zentrum Berlin f\"ur Materialien und Energie, D-14109 Berlin, Germany}

\author{Oleg I. Utesov}
\affiliation{Center for Theoretical Physics of Complex Systems, Institute for Basic Science, Daejeon 34126, Republic of Korea}

\author{Chen Luo}
\affiliation{Helmholtz-Zentrum Berlin f\"ur Materialien und Energie, D-14109 Berlin, Germany}

\author{Florin Radu}
\affiliation{Helmholtz-Zentrum Berlin f\"ur Materialien und Energie, D-14109 Berlin, Germany}

\author{Sebastian Wintz}
\affiliation{Helmholtz-Zentrum Berlin f\"ur Materialien und Energie, D-14109 Berlin, Germany}

\author{Markus Weigand}
\affiliation{Helmholtz-Zentrum Berlin f\"ur Materialien und Energie, D-14109 Berlin, Germany}

\author{Simone Finizio}
\affiliation{Swiss Light Source, Paul Scherrer Institute, 5232 Villigen PSI, Switzerland}
\affiliation{Helmholtz-Zentrum Berlin f\"ur Materialien und Energie, D-14109 Berlin, Germany}

\author{Moritz Winter}
\affiliation{Max Planck Institute for Chemical Physics of Solids, 01187, Dresden, Germany}
\affiliation{Dresden Center for Nanoanalysis, cfaed, Technical University Dresden, 01069 Dresden, Germany}

\author{Alexander Tahn}
\affiliation{Dresden Center for Nanoanalysis, cfaed, Technical University Dresden, 01069 Dresden, Germany}

\author{Bernd Rellinghaus}
\affiliation{Dresden Center for Nanoanalysis, cfaed, Technical University Dresden, 01069 Dresden, Germany}

\author{Kosuke Karube}
\affiliation{RIKEN Center for Emergent Matter Science (CEMS), Wako 351-0198, Japan}

\author{Yoshinori Tokura}
\affiliation{RIKEN Center for Emergent Matter Science (CEMS), Wako 351-0198, Japan}
\affiliation{Department of Applied Physics, University of Tokyo, Tokyo 113-8656, Japan}
\affiliation{Tokyo College, University of Tokyo, Tokyo 113-8656, Japan}

\author{Yasujiro Taguchi}
\affiliation{RIKEN Center for Emergent Matter Science (CEMS), Wako 351-0198, Japan}

\author{Jonathan S. White}
\affiliation{Laboratory for Neutron Scattering and Imaging (LNS), Paul Scherrer Institute (PSI), CH-5232 Villigen, Switzerland}


\date{\today}

\begin{abstract}

The cubic $\beta$-Mn-type alloy \czm \, is a chiral helimagnet that exhibits a peculiar temperature-dependent behavior in the spiral pitch, which decreases from 130\,nm at room temperature to 70\,nm below 20\,K. Notably, this shortening is also accompanied by a structural transition of the metastable skyrmion texture, transforming from a hexagonal lattice to a square lattice of elongated skyrmions. The underlying mechanism of these transformations remain unknown, with interactions potentially involved including temperature-dependent Dzyaloshinskii-Moriya interaction, magnetocrystalline anisotropy, and exchange anisotropy. Here, x-ray resonant magnetic small-angle scattering in vectorial magnetic fields was employed to investigate the temperature dependence of the anisotropic properties of the helical phase in \czm. Our results reveal quantitatively that the magnitude of the anisotropic exchange interaction increases by a factor of 4 on cooling from room temperature to 20\,K, leading to a 5\% variation in the helical pitch within the (001) plane at 20\,K. While anisotropic exchange interaction contributes to the shortening of the spiral pitch, its magnitude is insufficient to explain the variation in the spiral periodicity from room to low temperatures. Finally, we demonstrate that magnetocrystalline and exchange anisotropies compete, favoring different orientations of the helical vector in the ground state.
\end{abstract}

\maketitle

\section{Introduction}

In recent years, there has been significant interest in the investigation of particle-like magnetic solitons, primarily due to their potential use as fundamental building blocks for spintronic devices \cite{bogdanov1994thermodynamically,nagaosa2013topological,fert2017magnetic}. These intriguing textures include chiral solitons \cite{togawa2013interlayer}, and topologically-nontrivial spin hedgehogs \cite{kanazawa2011large} and skyrmions \cite{fert2017magnetic,muhlbauer2009skyrmion,nayak2017magnetic}, which are known to give rise to emergent electromagnetic phenomena essential for the advancement of dissipationless electronics and next-generation technologies \cite{nagaosa2013topological,fert2017magnetic}. A key aspect in comprehending and harnessing these magnetic structures lies in the exploration of their manifestation in noncentrosymmetric magnetic materials, which offer a fertile ground for the formation of various twisted spin textures \cite{tokura2020magnetic}.

\begin{figure*}
\begin{center}
\includegraphics[width=1\linewidth]{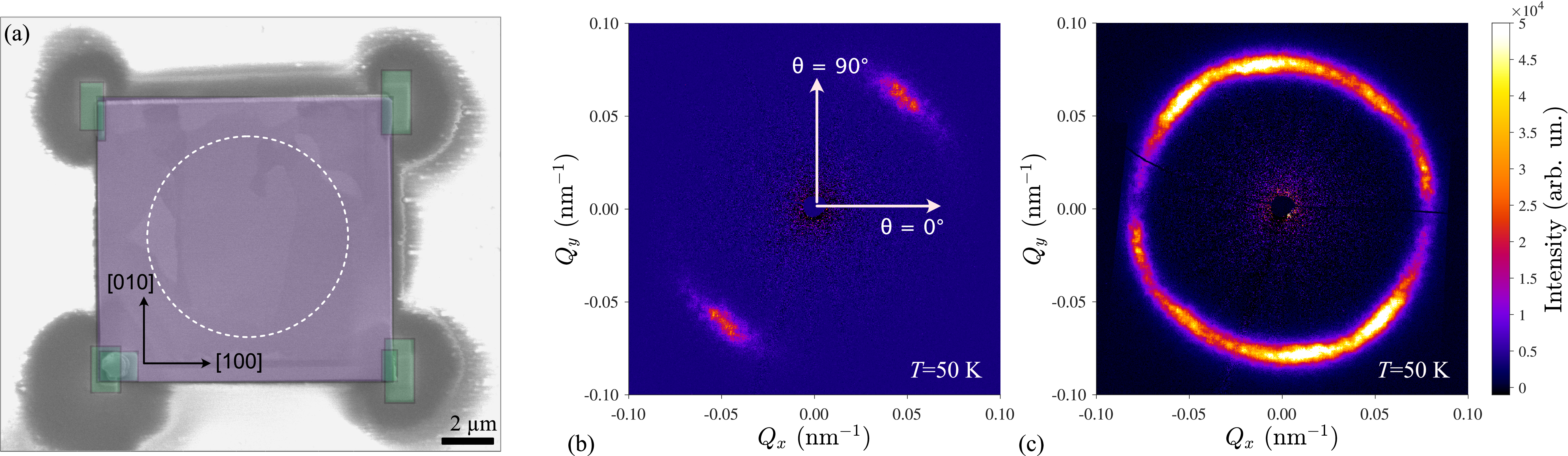}\vspace{3pt}
        \caption{SEM image of the (001) \czm~thin plate (purple pseudo colour) attached to a Si$_3$N$_4$ membrane by Pt contacts (green). The circular aperture drilled in the gold layer on the back side of the membrane is shown as a dashed line. (b) Zero-field REXS pattern measured at $T=50$\,K after the field training $\mu_0 H=0.12$\,T applied at $\theta=45^\circ$. (c) REXS patterns measured at 50\,K summed up over all azimuthal magnetic field training angles $\theta=0...180^\circ$ with a $3^\circ$ step.}
        \label{fig1}
\end{center}
\end{figure*}

In cubic chiral magnets, a rich variety of modulated magnetic states arises from the competition between several fundamental interactions, such as the Heisenberg exchange interaction, the antisymmetric Dzyaloshinskii-Moriya interaction (DMI), the anisotropic exchange interaction (AEI), and cubic anisotropy \cite{dzyaloshinsky1958thermodynamic,moriya1960anisotropic,bak1980theory,maleyev2006cubic,chacon2018observation,bannenberg2019multiple}. Recently, a new family of chiral cubic Co-Zn-Mn intermetallics with tunable magnetic properties has been discovered \cite{tokunaga2015new}. These materials crystallize in the $\beta$-Mn-type structure and exhibit rich magnetic phase diagrams that vary in their structure according to the precise composition \cite{karube2020metastable}. Moreover, in addition to the generally complex landscape of magnetic interactions, effects of magnetic frustration, and of both structural and magnetic disorder, also play crucial roles in the phase diagram of these compounds \cite{karube2020metastable,karube2016robust,karube2017skyrmion,karube2018disordered,yu2018transformation,nakajima2019correlation,bocarsly2019deciphering,ukleev2021frustration,preissinger2021vital,hicken2021megahertz,nagase2021observation,white2022small}.

Co$_8$Zn$_8$Mn$_4$ sustains particular attention because it hosts an equilibrium skyrmion lattice phase at room temperature under moderate magnetic fields \cite{karube2016robust}. Intriguingly, for \czm~as well as for similar compounds with different Mn concentrations, the helical spiral period $\lambda$ is observed to decrease significantly as the temperature decreases, leading to a substantial increase in the spiral wavevector $Q_0=2\pi/\lambda \sim D/J$, where $D$ is the DMI constant and $J$ is the exchange stiffness \cite{karube2020metastable}. In \czm~for example, the increase of the helical wavevector by $\sim~50$~\% on cooling from 300\,K to below 20\,K affects the structural properties of either equilibrium helical or metastable skyrmion lattice states, and plays a key role in the transition between different high- and low-temperature metastable skyrmion lattice coordinations \cite{karube2016robust,morikawa2017deformation,ukleev2019element,henderson2022skyrmion}.

\begin{figure*}
\begin{center}
\includegraphics[width=1\linewidth]{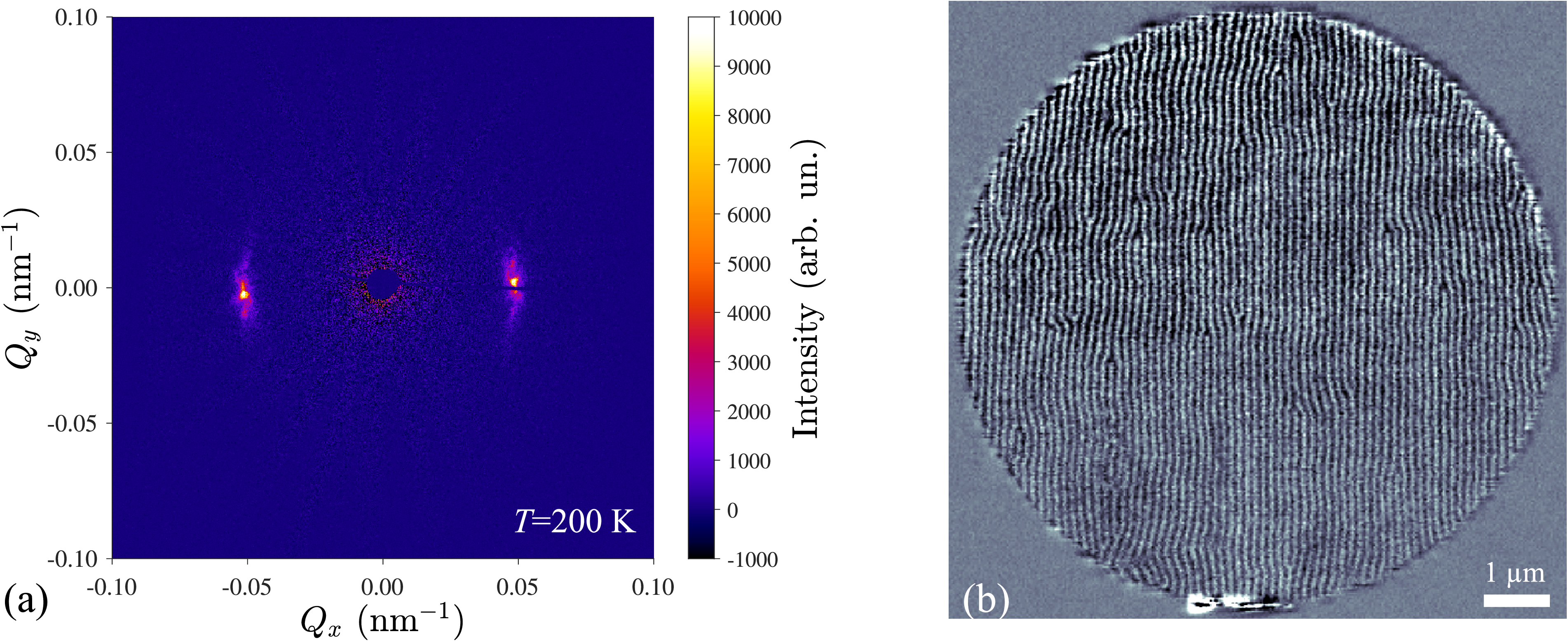}\vspace{3pt}
        \caption{(a) Zero-field REXS pattern measured at $T=200$\,K after the field training $\mu_0 H=0.12$\,T applied at $\theta=0^\circ$. (b) SXTM image of the aligned helical texture at $T=200$\,K obtained using circularly polarized x-rays, which are sensitive to the out-of-plane component of magnetization.}
        \label{fig2}
\end{center}
\end{figure*}

The role of anisotropic interactions, including cubic and exchange anisotropies, has been extensively studied in the context of how they determine the properties of chiral magnets  \cite{chacon2018observation,bannenberg2019multiple,qian2018new,leonov2023reorientation}. Specifically, the competition between these two interactions significantly affects macroscopic parameters such as the transition magnetic fields between different magnetic states \cite{maleyev2006cubic,grigoriev2015spiral,leonov2023reorientation}. However, disentangling their contributions through direct experimental observations is challenging, since both interactions can simultaneously influence the macroscopic observables. Even neutron scattering techniques, which provide microscopic information about materials, often require additional theoretical models to distinguish between the contributions of cubic and exchange anisotropies. Therefore, a comprehensive investigation of the interplay between these interactions in chiral magnets is needed for a complete understanding of their magnetic phase diagrams and how to exploit these exotic magnetic phases for applications.

To gain a deeper understanding of the microscopic magnetic interactions in \czm, it is essential to probe them independently, without relying on indirectly related parameters. Recently, the temperature dependence of both the magnetocrystalline anisotropies \cite{preissinger2021vital} and the exchange stiffness \cite{ukleev2022spin} have been quantified by means of ferromagnetic resonance and neutron spectroscopy, respectively. From the latter study, it was suggested that besides an increase in the exchange stiffness at low temperature, a quantitatively accurate description of the size of the observed spiral wavevector elongation requires a further concomitant variation on cooling of other interactions involved, such as an increase in the DMI constant.

In this study, we report direct measurements of the anisotropic exchange interaction (AEI) in \czm~using resonant elastic small-angle x-ray scattering \cite{ukleev2021signature}. Through the investigation of the temperature dependence of the AEI constant, we successfully parameterise a further fundamental interaction that participates the unusual temperature variation of the helical wavevector and the transformation of the metastable skyrmion lattice. We discuss the results within the context of composition-tuning of competing anisotropies in Co-Zn-Mn compounds as means to gain control over the observable helical structures in these materials.

\section{Experimental}

The single crystal of \czm~used for resonant elastic soft x-ray scattering (REXS) experiments was grown using the Bridgman method, as described in Ref. \cite{karube2016robust}. From the oriented bulk specimen, a \czm~lamella with a thickness of approximately $190$\,nm and an area of approximately $10\times10$\,$\mu$m$^2$ was cut parallel to the (100) plane and subsequently thinned using a focused ion beam system (FIB). To facilitate the experiments, the lamella was attached to a gold-coated Si$_3$N$_4$ membrane (Silson Ltd, UK) using four Pt contacts deposited by FIB. Before attaching the lamella, a circular aperture with a 8\,$\mu$m diameter was milled through the membrane and the gold layer to allow x-ray transmission through the sample. A scanning electron microscopy (SEM) image of the sample is shown in Fig. \ref{fig1}a. For more comprehensive details about the sample fabrication routine, we refer to Ref. \cite{ukleev2019element}. More details on the sample synthesis, characterisation, and nanofabrication are given in the Supplemental Material \cite{supp}.

The \czm~sample, mounted on the membrane, was placed onto a Cu sample holder and loaded into the vector-field cryomagnet VEKMAG at the PM-2 beamline \cite{noll2016mechanics} at BESSY II, Helmholtz-Zentrum Berlin, Germany. Note that the in-plane orientation of the sample with respect to the laboratory frame of reference was chosen arbitrarily.

Circularly polarized soft x-ray beams were utilized at an energy tuned to the Co $L_3$ edge ($E=775.8$\,eV), where the intensity of magnetic satellites reached its maximum. The sample was then cooled down in absence of a magnetic field to a temperature of 20\,K, which is a few degrees above the spin-glass transition \cite{karube2020metastable,ukleev2021frustration}. The details of the small-angle REXS experiment in a vector field closely matched those reported in Refs. \cite{ukleev2021signature,baral2023direct}. A charge-coupled device (CCD) detector ALEX-i 4k4k with $4096\times4096$ pixels$^2$ (GreatEyes GmbH, Berlin, Germany) was employed, allowing an accessible range of scattering angles $2\theta$ of $\pm 3^\circ$. Since the detector pixel size (15\,$\mu$m$\times$15\,$\mu$m) did not limit the resolution in this particular experiment, a hardware binning of 2 was utilized for faster image acquisition. The typical acquisition time for each small-angle scattering pattern was 30\,seconds.

\section{Results and discussion}

\begin{figure*}
\begin{center}
\includegraphics[width=1\linewidth]{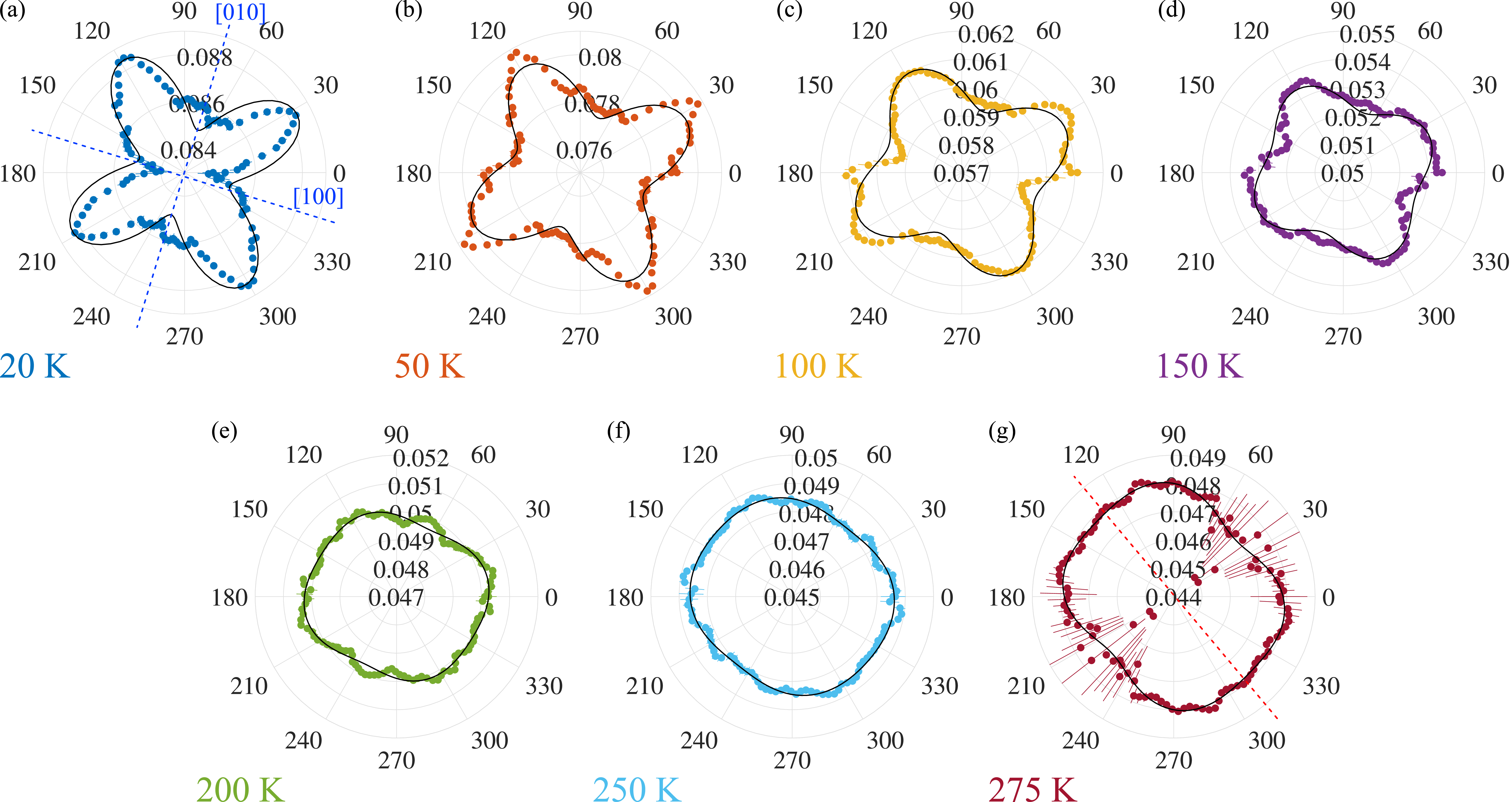}\vspace{3pt}
        \caption{Polar plots of the helical wavevector magnitude $Q$ in nm$^{-1}$ units as a function of azimuthal angle $\theta$ at (a) 20\,K, (b) 50\,K, (c) 100\,K, (d) 150\,K (e) 200\,K, (f) 250\,K, (g) 275\,K. The data is symmetrized along the horizontal axis. Solid lines correspond to the fit according to Eq. \ref{eq1}. The dashed line in panel (g) corresponds to the direction of the uniaxial anisotropy.}
        \label{fig3}
\end{center}
\end{figure*}

After zero-field cooling the helical propagation vector in the (001) plane of the \czm~crystal was determined using a vector magnetic field training procedure. This procedure involves saturating the sample with a vectorial magnetic field of 150\,mT applied at an angle $\theta$ in the sample plane. Here, $\theta=0^\circ$ corresponds to the vertical direction $\mathbf{z}$ in the laboratory frame of reference. Subsequently, the magnetic field was gradually reduced to zero. Similar to the case of FeGe \cite{ukleev2021signature}, this training procedure results in the alignment of the spiral propagation vector at any $\theta$ angle in the sample plane (see Fig. \ref{fig1}b). To map out the period of the spiral as a function of the propagation direction in the (001) plane, measurements were carried out with $\theta$ steps of $3^\circ$, covering a temperature range from 20\,K to 275\,K. Two-dimensional REXS patterns obtained at 50\,K by summing up measurements over all azimuthal magnetic field training angles ($\theta=0...180^\circ$) with a $3^\circ$ step are shown in Fig. \ref{fig1}c. The field-training procedure was working successfully for all measured temperatures and magnetic field angles in the (001) plane. 

As a demonstration of the field training procedure, we show a reciprocal (Fig. \ref{fig2}a) and real space image (Fig. \ref{fig2}b) of the resulting magnetic helical structure, measured at 200\,K. At this temperature the magnitude of the spiral wavevector is approximately 1.5 times larger than at 50\,K. The real-space image of the aligned helical texture in the same sample was obtained using scanning transmission x-ray microscopy (STXM) at the Co $L_3$ edge with the MAXYMUS instrument at BESSY II \cite{nolle2012note} (Fig. \ref{fig2}b), operated by the Helmholtz-Zentrum f\"ur Materialien und Energie.

The REXS patterns obtained for each temperature and azimuthal magnetic field training angle $\theta$ show a pair of Bragg peaks corresponding to the helical texture aligned in the corresponding in-plane direction. In Figs. \ref{fig3}a-g, we present the temperature dependence of the extracted peak positions $Q(\theta)$. At low temperatures ($T\leq150$\,K), the dependence clearly exhibits a four-fold character, as expected for the cubic symmetry in the (001) plane \cite{maleyev2006cubic}.

The spiral wavevector $Q$ in the (001) plane reaches a local maximum (corresponding to a minimum real-space period $\lambda$) when aligned along $[110]$-equivalent crystal axes and a minimum when aligned along $[100]$ directions. At 20\,K, the magnitude of the $Q$-vector oscillation in the (001) plane reaches approximately 5\%. This general behavior of the helical wavevector directly provides the sign of the AEI constant $F$ to be negative, similar to both MnSi \cite{grigoriev2009crystal} and Cu$_2$OSeO$_3$ \cite{baral2023direct,moody2021experimental}, which in principle is expected to favor a preferred alignment of helical wavevectors along the body-diagonals of the cubic lattice $\langle111\rangle$ \cite{maleyev2006cubic}. \czm~is known to display a preferred alignment of the helical wavevectors with the cubic $\langle 100 \rangle$ axes at all temperatures~\cite{karube2020metastable,karube2016robust}, consistent in principle with a positive value of $F$. The negative sign of $F$ found experimentally here therefore implies a competition between the different helical wavevector alignment tendencies promoted by a predominant cubic anisotropy \cite{preissinger2021vital} favoring the observed $\langle 100 \rangle$ directions, and the weaker AEI. A similar situation was recently observed in Cu$_2$OSeO$_3$~\cite{baral2023direct}, where the interplay of frustrated anisotropies stabilizes exotic tilted conical and disordered skyrmion phases at low temperatures when a magnetic field is applied along [001]. 

\begin{figure*}
\begin{center}
\includegraphics[width=1\linewidth]{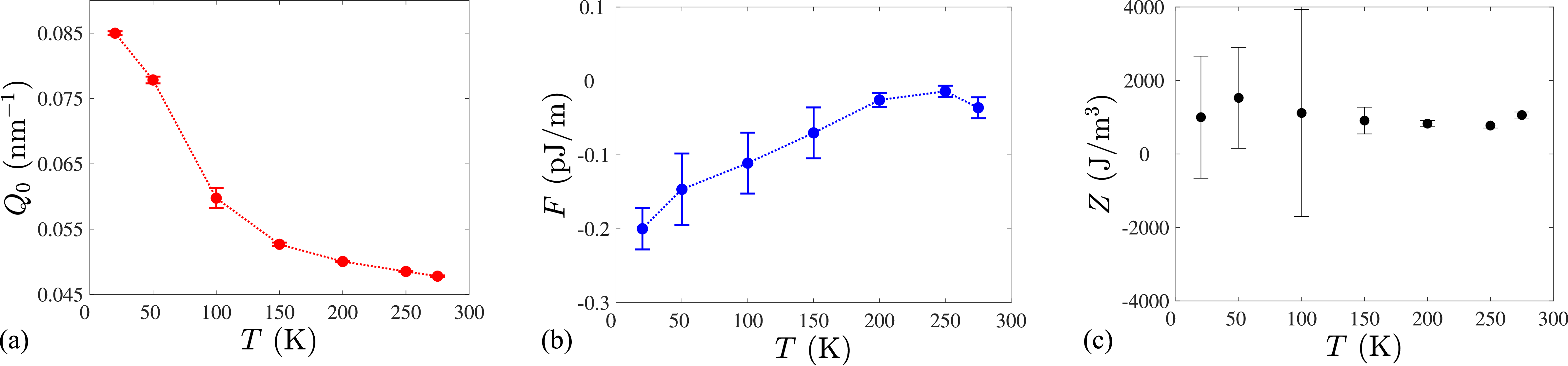}\vspace{3pt}
        \caption{Temperature dependence of (a) spiral wavevector $Q_0$, (b) the exchange anisotropy constant $F$, and (c) the strain-induced uniaxial anisotropy constant $Z$ extracted from the fit of $Q(\theta)$ according to Eq. \ref{eq1}.}
        \label{fig4}
\end{center}
\end{figure*}

Upon elevating the temperature above 20~K, the $Q(\theta)$ dependence becomes more isotropic (Fig. \ref{fig3}). At around $T\sim200$\,K and above, the four-fold feature is no longer pronounced, but the distribution is not completely isotropic either. Instead, a weak uniaxial distortion of $Q(\theta)$ is observed that can be successfully explained by considering an effect of spiral distortion induced by uniaxial tensile strain \cite{ukleev2020metastable,baral2023direct}. To model the $Q(\theta)$ dependence for all temperatures, we applied the following equation, which takes into account both the AEI and uniaxial strain-induced anisotropy  \cite{baral2023direct}:

\begin{eqnarray} \label{eq1}
Q  &=& Q_0 \Biggl\{  1 - \dfrac{F\sin^{2}{2\theta}}{4J}\Biggr\} - \dfrac{JZ^{2}}{8D^{3}}\sin^{4}(\theta-\phi). 
\end{eqnarray}

Here $Z$ represents the uniaxial strain-induced anisotropy constant, and $\phi$ indicates the corresponding strain-axis direction. The model is similar to the one used for Cu$_2$OSeO$_3$ \cite{baral2023direct}, with the exception that a term containing the conical angle of the spiral is 0 since the present $Q(\theta)$ measurements were carried out at zero magnetic field. By applying the equation~\ref{eq1} to fit the $Q(\theta)$ dependent data, we account for the effects of an observed uniaxial distortion and explain its interplay with the AEI in the \czm\,lamella.

The solid lines in Figure \ref{fig3}a-g represent the $Q(\theta)$ curves fitted according to Eq. \ref{eq1}. In the fitting procedure, the measured ($T>70$\,K) or extrapolated according to the power law ($T<70$\,K) exchange $J$ and $D$ constants at each temperature are taken from a previous neutron scattering experiment \cite{ukleev2022spin}, and the parameters $Q_0$, $F$, $Z$, and $\phi$ adjusted to obtain the best fit to the data. An offset of $\theta$ and $\phi$ angles was introduced in the fit to account for the arbitrary sample orientation with respect to the laboratory frame of reference. Remarkably, this simple model including contributions from both the AEI and uniaxial anisotropy, provides an excellent agreement with the data and allows for the quantitative extraction of the anisotropic parameters (Fig. \ref{fig4}(b)-(c)). Additional terms including those that are second-order in the AEI and the cubic anisotropy do not improve the fit and provide the same fitting results within the experimental error bars.

The temperature dependence of $Q_0$ obtained from the fitting results shown in Figure \ref{fig4}a agrees well with previous reports from neutron and x-ray scattering experiments \cite{karube2016robust,ukleev2020metastable,ukleev2022spin,henderson2022skyrmion}. The most important result is shown in Figure \ref{fig4}b, where $F$ is fitted to be negative throughout the entire temperature range and tends to monotonically increase in magnitude towards lower temperatures. From $F=-0.0036\pm0.0015$ pJm$^{-1}$ at $T=275$\,K the AEI constant reaches a value of $F=-0.20\pm0.02$ pJm$^{-1}$ at 20\,K. Fig. \ref{fig4}c shows that the fitted uniaxial anisotropy remains nearly constant across the whole temperature range (Fig. \ref{fig4}c), and its contribution to the $Q$-vector magnitude is insignificant at low temperature as compared to the AEI. We mention nevertheless, that the, the presence of the strain-induced anisotropy offers the possibility for stabilizing a chiral soliton lattice in \czm~close to room temperature \cite{ukleev2020metastable,brearton2023observation}.

Our findings provide valuable microscopic insights into the competition between anisotropies in \czm~and shed light on their role in determining the magnetic properties, including the unusual temperature variation of the helical wavevector and the transformation of the metastable skyrmion lattice state. Notably, $|F|$ increases by a factor of four on cooling from 275\,K to 20\,K, yet this variation alone cannot explain the dramatic change of $Q_0$ length on cooling, nor the persistent helical wavevector alignment with the cubic $\langle 100 \rangle$ directions. Instead, the significant temperature dependences of $J$ and $D$ ~\cite{ukleev2022spin}, in combination with the cubic anisotropy \cite{preissinger2021vital} dominate the observed fundamental behavior of the helical wavevectors and the shape change of the metastable skyrmion lattice \cite{karube2016robust,morikawa2017deformation} in \czm, with the underlying AEI playing a minor role. Nonetheless, we note that the AEI can be brought to bear on the observable properties through a suppression of cubic anisotropy so that it is the AEI that determines the helical wavevector alignment. In Co-Zn-Mn compounds this can be achieved by reducing the Mn content \cite{preissinger2021vital}. Indeed, the preferred alignment of the helical wavevectors does eventually change to be along the $\langle111\rangle$ directions in Mn-free Co$_{10}$Zn$_{10}$ \cite{karube2020metastable}. This suggests that compositional tuning can provide control of the helical wavevector properties in very low-Mn content Co-Zn-Mn compounds, through fine-tuning the balance between cubic anisotropy and AEI.

\begin{figure}
\includegraphics[width=1\linewidth]{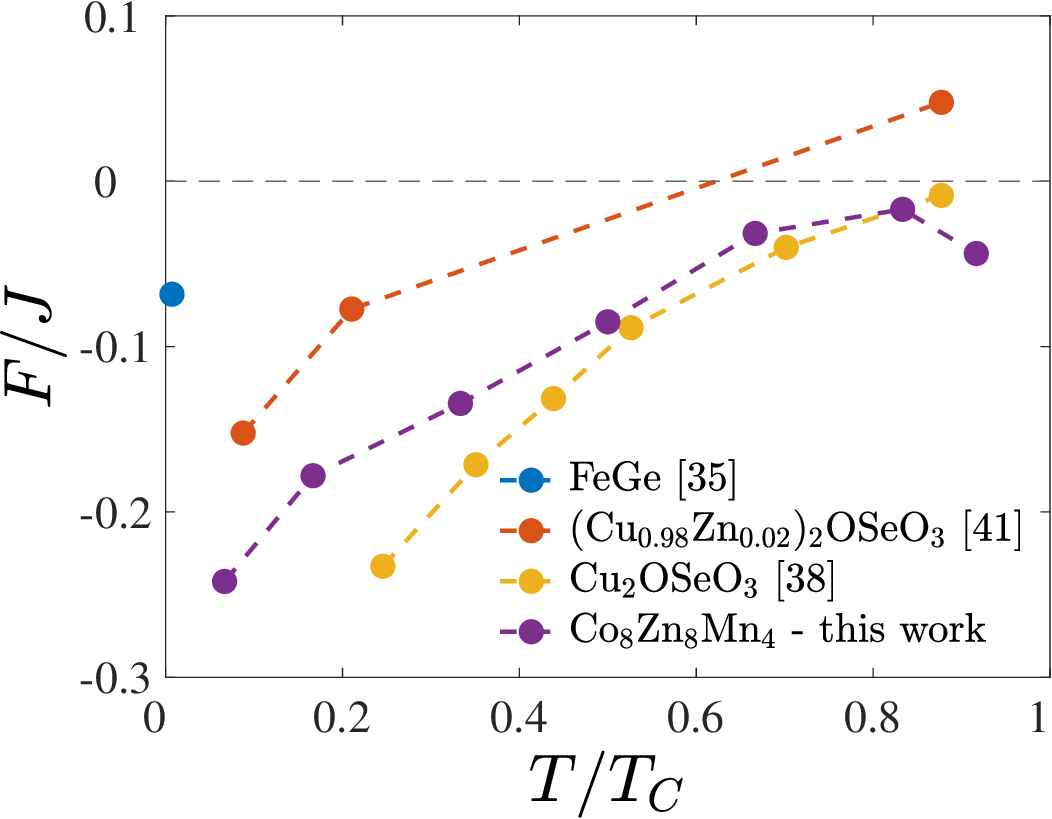}\vspace{3pt}
        \caption{Comparison between the $F/J$ ratios vs. reduced temperatures  for FeGe \cite{ukleev2021signature}, Zn-doped Cu$_2$OSeO$_3$ \cite{moody2021experimental}, pristine Cu$_2$OSeO$_3$ \cite{baral2023direct} and \czm~(this work).}
        \label{fig5}
\end{figure}

In Fig. \ref{fig5}, we compare the extracted $F/J$ ratio vs. reduced temperature $T/T_\textrm{C}$ for the cubic chiral magnets FeGe \cite{ukleev2021signature}, (Cu$_{0.98}$Zn$_{0.02}$)$_2$OSeO$_3$ \cite{moody2021experimental}, Cu$_2$OSeO$_3$ \cite{baral2023direct} and \czm~measured experimentally using neutron or x-ray scattering probes. Importantly, the magnitude of the AEI at low temperatures in \czm~is even larger than that in Cu$_2$OSeO$_3$ \cite{baral2023direct}. This raises interesting questions concerning why tilted conical and thermal-equilibrium low-temperature skyrmion phases are apparently absent in bulk \czm~in contrast to Cu$_2$OSeO$_3$ where they are readily observed \cite{qian2018new,chacon2018observation}. Understanding the interplay between the relatively large AEI and the cubic magnetocrystalline anisotropy in \czm, as well as their relative size, compared with those found in other chiral magnets is important in terms of clarifying their effect on the stability of different chiral magnetic phases, and requires further theoretical investigations. 

The microscopic parameters derived in the present study, along with previous works on the cubic anisotropy \cite{preissinger2021vital} and exchange stiffness \cite{ukleev2022spin}, provide a basis for theoretical models to elucidate the complex magnetic phase diagrams and the stability of exotic magnetic states in \czm~at different temperatures and magnetic fields. The Co-Zn-Mn class of chiral magnets challenges the time-honoured Bak-Jensen model and calls for inclusion of ingredients that take into account the microscopic richness of these materials, such as short-range magnetic order and frustration-driven fluctuations typical to $\beta$-Mn alloys \cite{eriksson2005magnetic,paddison2013emergent,ukleev2021frustration,yamauchi2020high,white2022small}. Such theoretical studies could offer valuable insights into the underlying mechanisms governing the magnetic properties over multiple magnetic length-scales in this class of chiral skyrmion hosts.

\section{Conclusion}

In summary, resonant x-ray scattering techniques have been employed to provide new microscopic measurements of the anisotropic exchange interaction (AEI) in the chiral cubic magnet \czm. Applying a theoretical model that incorporates both the AEI and uniaxial strain-induced anisotropy contributions to the data allows for the quantitative extraction of the AEI parameter $F$. The observation of a negative value for $F$ over the entire temperature range below $T_\textrm{C}$ is at odds with the positive value expected according to the known preferred alignment of the helical wavevector, which itself indicates the existence of an underlying competition between the AEI and a more dominant magnetocrystalline anisotropy on the properties of the helical order \czm. 

Interestingly, despite the comparable magnitude of $F$ to that of Cu$_2$OSeO$_3$, the AEI does not seem to enrich the magnetic phase diagram of \czm~with tilted conical and disordered skyrmion phases. Additionally, short-range magnetic order and magnetic fluctuations, which drive the disordered skyrmion phase in the highly frustrated Co$_7$Zn$_7$Mn$_6$, do not stabilize low-temperature skyrmions in \czm~either \cite{ukleev2021frustration}. However, the chemical tunability of Co-Zn-Mn compounds offers an exciting opportunity for further engineering of magnetic anisotropies and helical structures, potentially enabling the tuning of the skyrmion phase diagram and inducing novel twisted phases in this class of materials.

Overall, this study contributes to the deeper quantitative understanding of the magnetic interactions in \czm, paving the way for future research in this area and opening up possibilities for exploring new magnetic phases and functionalities in chiral cubic magnets through controlled manipulation of their magnetic anisotropy.

\section*{Acknowledgements}

Authors thank E. Deckardt, M. Bednarzik and Th. Jung for their help in preparation of the membranes at PSI. The REXS experiment was carried out at the beamline PM-2 VEKMAG at BESSY II synchrotron as a part of the proposal 222-11296-ST. We thank the Helmholtz-Zentrum Berlin f\"ur Materialien und Energie for the allocation of synchortron radiation beamtime at both VEKMAG and MAXYMUS. Authors acknowledge funding from JST-CREST (Grant No. JPMJCR20T1). We also acknowledge financial support for the VEKMAG project and for the PM2-VEKMAG beamline by the German Federal Ministry for Education and Research (BMBF 05K2010, 05K2013, 05K2016, 05K2019) and by HZB. F.R. acknowledges funding by the German Research Foundation via Project No. SPP2137/RA 3570. O. I. U. acknowledges support from the Institute for Basic Science (IBS) in the Republic of Korea through Project No. IBS-R024-D1. M.W. acknowledges support from the International Max Planck Research School for Chemistry and Physics of Quantum Materials (IMPRS-CPQM), and B.R. is grateful for funding from the DFG through SPP 2137, project no.\ 403503416. J.S.W. acknowledges funding from the SNSF Project 200021\_188707.

\end{document}